\newcommand{\CLASSINPUTtoptextmargin}{19mm}%
\newcommand{\CLASSINPUTbottomtextmargin}{43mm}%
\newcommand{\CLASSINPUTinnersidemargin}{12.9mm}%
\newcommand{\CLASSINPUToutersidemargin}{12.9mm}%
\documentclass[conference,10pt,a4paper]{IEEEtran}
\usepackage{amsmath}
\usepackage{times}
\usepackage{graphicx}
\usepackage{multirow}
\usepackage[none]{hyphenat}
\usepackage{float}
\usepackage{subfig}
\usepackage{iftex}
\usepackage{xurl}
\usepackage{caption}
\usepackage{subcaption}
\usepackage{subcaption}
\usepackage{booktabs}
\ifPDFTeX%
\usepackage{t1enc}
\usepackage{times}
\fi%
\ifLuaTeX%
\usepackage{fontspec}
\fi%
\makeatletter

\def\@maketitle{\newpage
\bgroup\par\addvspace{0.5\baselineskip}\centering%
\ifCLASSOPTIONtechnote
   {\bfseries\large\@IEEEcompsoconly{\sffamily}\@title\par}\vskip 1.3em{\lineskip .5em\@IEEEcompsoconly{\sffamily}\@author
   \@IEEEspecialpapernotice\par{\@IEEEcompsoconly{\vskip 1.5em\relax
   \@IEEEtitleabstractindextextbox{\@IEEEtitleabstractindextext}\par
   \hfill\@IEEEcompsocdiamondline\hfill\hbox{}\par}}}\relax
\else
   \vskip0.2em{\EuMWtitlesize\ifCLASSOPTIONtransmag\bfseries\LARGE\fi\@IEEEcompsoconly{\sffamily}\@IEEEcompsocconfonly{\normalfont\normalsize\vskip 2\@IEEEnormalsizeunitybaselineskip
   \bfseries\Large}\@title\par}\vskip1.0em\par
   \ifCLASSOPTIONconference%
      {\@IEEEspecialpapernotice\mbox{}\vskip\@IEEEauthorblockconfadjspace%
       \mbox{}\hfill\begin{@IEEEauthorhalign}\@author\end{@IEEEauthorhalign}\hfill\mbox{}\par}\relax
   \else
      \ifCLASSOPTIONpeerreviewca
         {\@IEEEcompsoconly{\sffamily}\@IEEEspecialpapernotice\mbox{}\vskip\@IEEEauthorblockconfadjspace%
          \mbox{}\hfill\begin{@IEEEauthorhalign}\@author\end{@IEEEauthorhalign}\hfill\mbox{}\par
          {\@IEEEcompsoconly{\vskip 1.5em\relax
           \@IEEEtitleabstractindextextbox{\@IEEEtitleabstractindextext}\par\hfill
           \@IEEEcompsocdiamondline\hfill\hbox{}\par}}}\relax
      \else
         \ifCLASSOPTIONtransmag
           {\@IEEEspecialpapernotice\mbox{}\vskip\@IEEEauthorblockconfadjspace%
            \mbox{}\hfill\begin{@IEEEauthorhalign}\@author\end{@IEEEauthorhalign}\hfill\mbox{}\par
           {\vspace{0.5\baselineskip}\relax\@IEEEtitleabstractindextextbox{\@IEEEtitleabstractindextext}\vspace{-1\baselineskip}\par}}\relax
         \else
           {\lineskip.5em\@IEEEcompsoconly{\sffamily}\sublargesize\@author\@IEEEspecialpapernotice\par
           {\@IEEEcompsoconly{\vskip 1.5em\relax
            \@IEEEtitleabstractindextextbox{\@IEEEtitleabstractindextext}\par\hfill
            \@IEEEcompsocdiamondline\hfill\hbox{}\par}}}\relax
         \fi
      \fi
   \fi
\fi\par\addvspace{0.0\baselineskip}\egroup}

\def\EuMWtitlesize{\@setfontsize{\EuMWtitlesize}{24}{24pt}}
\def\EuMWauthorsize{\@setfontsize{\EuMWauthorsize}{11}{11pt}}
\def\EuMWaffilsize{\@setfontsize{\EuMWaffilsize}{10}{10pt}}
\def\EuMWcaptionsize{\@setfontsize{\EuMWcaptionsize}{9}{10pt}}
\def\EuMWbibsize{\@setfontsize{\EuMWbibsize}{8}{10pt}}

\def\@IEEEauthorblockNstyle{\EuMWauthorsize\@IEEEcompsocnotconfonly{\sffamily}\@IEEEcompsocconfonly{\large}}
\def\@IEEEauthorblockAstyle{\EuMWaffilsize\@IEEEcompsocnotconfonly{\sffamily}\@IEEEcompsocconfonly{\itshape}\@IEEEcompsocconfonly{\large}}
\def\@IEEEauthordefaulttextstyle{\EuMWauthorsize\@IEEEcompsocnotconfonly{\sffamily}\sublargesize}

\def\thebibliography#1{\section*{\refname}%
    \addcontentsline{toc}{section}{\refname}%
    \EuMWbibsize\@IEEEcompsocconfonly{\small}\vskip 0.3\baselineskip plus 0.1\baselineskip minus 0.1\baselineskip
    \list{\@biblabel{\@arabic\c@enumiv}}%
    {\settowidth\labelwidth{\@biblabel{#1}}%
    \leftmargin\labelwidth
    \advance\leftmargin\labelsep\relax
    \itemsep \IEEEbibitemsep\relax
    \usecounter{enumiv}%
    \let\p@enumiv\@empty
    \renewcommand\theenumiv{\@arabic\c@enumiv}}%
    \let\@IEEElatexbibitem\bibitem%
    \def\bibitem{\@IEEEbibitemprefix\@IEEElatexbibitem}%
\def\newblock{\hskip .11em plus .33em minus .07em}%
\ifCLASSOPTIONtechnote\sloppy\clubpenalty4000\widowpenalty4000\interlinepenalty100%
\else\sloppy\clubpenalty4000\widowpenalty4000\interlinepenalty500\fi%
    \sfcode`\.=1000\relax}

\long\def\@makecaption#1#2{%
\ifx\@captype\@IEEEtablestring%
\par\@IEEEtabletopskipstrut
\else
\@IEEEfigurecaptionsepspace
\fi
\setbox\@tempboxa\hbox{\normalfont\footnotesize {#1.}\nobreakspace\nobreakspace #2}%
\ifdim \wd\@tempboxa >\hsize%
\setbox\@tempboxa\hbox{\normalfont\footnotesize {#1.}\nobreakspace\nobreakspace}%
\parbox[t]{\hsize}{\normalfont\footnotesize\noindent\unhbox\@tempboxa#2}%
\else
\ifCLASSOPTIONconference \hbox to\hsize{\normalfont\footnotesize\hfil\box\@tempboxa\hfil}%
\else \hbox to\hsize{\normalfont\footnotesize\box\@tempboxa\hfil}%
\fi\fi
\ifx\@captype\@IEEEtablestring%
\@IEEEtablecaptionsepspace
\else
\fi}

\newlength\tablecaptiontotableskip
\newlength\figuretocaptionskip
\def\@IEEEfigurecaptionsepspace{\vskip\figuretocaptionskip\relax}%
\def\@IEEEtablecaptionsepspace{\vskip\tablecaptiontotableskip\relax}%

\def\abstract{\normalfont%
\@IEEEabskeysecsize\bfseries\textit{\abstractname}\,\bfseries\textit{---}\,%
\@IEEEgobbleleadPARNLSP}%

\def\IEEEkeywords{\normalfont%
\@IEEEabskeysecsize\bfseries\textit{\IEEEkeywordsname}\,\bfseries\textit{---}\,%
\@IEEEgobbleleadPARNLSP}%
\def\endIEEEkeywords{\relax\vspace{0.67ex}%
\par\if@twocolumn\else\endquotation\fi%
\normalsize\normalfont}%

\DeclareRobustCommand*{\EuMWauthorrefmark}[1]{\raisebox{0pt}[0pt][0pt]{\textsuperscript{#1}}}%
\def\@IEEEauthorblockNtopspace{0ex}
\def\@IEEEauthorblockAtopspace{1mm}
\def\tablename{Table}
\def\thetable{\arabic{table}}
\def\IEEEkeywordsname{Keywords}
\def\subsubsection{\@startsection{subsubsection}{3}{\z@}{1.5ex plus 1.5ex minus 0.5ex}%
{0.7ex plus .5ex minus 0ex}{\normalfont\normalsize\itshape}}%
\newlength{\CPheadmatchindent}%
\def\@seccntformat#1{\hbox to\CPheadmatchindent{\csname the#1dis\endcsname}\hskip 0.1em \relax}
\IEEEilabelindentA \parindent
\IEEEilabelindent \IEEEilabelindentA
\IEEEelabelindent \parindent
\IEEEdlabelindent \parindent
\IEEElabelindent \parindent
\makeatother

\begin{document}
\raggedbottom
%
%
%
\title{Digital Predistortion for Flux Control of Tunable Superconducting Qubits}
%

\author{%
\IEEEauthorblockN{%
Dharun Venkateswaran\EuMWauthorrefmark{1, 2$\dagger$}, 
Felice Francesco Tafuri\EuMWauthorrefmark{1$\ddagger$}, 
Yuanzheng Paul Tan\EuMWauthorrefmark{3$\star$}, 
Bruno Aznar Martinez\EuMWauthorrefmark{1},
Alisa Danilenko\EuMWauthorrefmark{1}, \\
Likai Yang\EuMWauthorrefmark{1},
Arnaud Carignan-Dugas\EuMWauthorrefmark{1}, 
Christoph Hufnagel\EuMWauthorrefmark{4}, 
Rainer Dumke\EuMWauthorrefmark{3, 4}, 
Philip Krantz\EuMWauthorrefmark{1},
Eric T. Holland\EuMWauthorrefmark{1}
}
\IEEEauthorblockA{%
\EuMWauthorrefmark{1}Quantum Engineering Solutions (QES), Keysight Technologies Inc., 1400 Fountaingrove Pkwy, Santa Rosa, CA 95403, USA\\
\EuMWauthorrefmark{2}Abdus Salam Centre for Theoretical Physics, Imperial College London, UK\\
\EuMWauthorrefmark{3}Division of Physics and Applied Physics, Nanyang Technological University, Singapore\\
\EuMWauthorrefmark{4}Centre for Quantum Technologies, National University of Singapore, Singapore \\
\EuMWauthorrefmark{$\dagger$}dv24@ic.ac.uk,
\EuMWauthorrefmark{$\ddagger$}francesco.tafuri@keysight.com,
\EuMWauthorrefmark{$\star$}ytan127@e.ntu.edu.sg
}
}
%
\maketitle

%
%
%
\begin{abstract}

Flux-tunable superconducting qubits rely on fast flux control pulses to implement two-qubit entangling quantum gates, a key building block for quantum algorithms. However, distortion effects introduced by non-ideal control electronics, parasitic components, and the cryogenic quantum chip response can all degrade the gate fidelity. We present a digital predistortion (DPD) framework for characterizing and then compensating for these distortions using a combination of infinite impulse response (IIR) and finite impulse response (FIR) filters. Experiments on a flux-tunable quantum processing unit (QPU) demonstrate a successful correction of step-response distortions on the flux-control line, with a compensated control signal showing only sub-percent deviations from the ideal target linear behavior. The demonstrated method enables automated rapid calibration of flux control channels for superconducting QPUs.

\end{abstract}
%
%

\section{Introduction}

Superconducting qubits have proven to be one of the most promising technologies for realizing large-scale QPUs \cite{QEng}, \cite{SQs-State-of-play}. The transmon qubit is the current industry standard for superconducting qubit design, thanks to its simplicity, robust fabrication, and scalability in both 2D and 3D architectures. To realize quantum algorithms on superconducting qubits, quantum gates are implemented as a sequence of microwave and baseband (BB) pulses sent along the XY and Z control lines, respectively, where the manipulation of the qubit-state across the X, Y, and Z axes is typically represented on a Bloch sphere \cite{QEng}. The transmon design also allows for the realization of flux-tunable qubits, whose transition frequency $(f_q)$ is adjustable by tuning the Josephson energy $(E_J)$ of a superconducting quantum interference device (SQUID) loop, by threading it with a magnetic flux $(\Phi)$. In a flux-tunable qubit architecture, two-qubit gates can be implemented using the controlled-Z (CZ) gate, where a combination of DC and BB waveforms are sent along the flux line \cite{QEng}, \cite{Rol_2019}. Hence, accurate flux pulse control is the basis for implementing high-fidelity two-qubit gates. Modern QPUs also deploy tunable couplers to achieve even faster two-qubit gates \cite{Yan_2018}. Similar digital predistortion techniques to the one presented in this paper have also been investigated to improve the control of tunable couplers \cite{DPD-TC}, as well as XY qubit control \cite{S_Gustav}. This paper, however, focuses on the compensation of distortion effects of the flux-control line.

\section{Distortion Effects in Flux-Tunable Qubits}

\begin{figure}[t]
    \centering
    \includegraphics[width=0.80\linewidth]{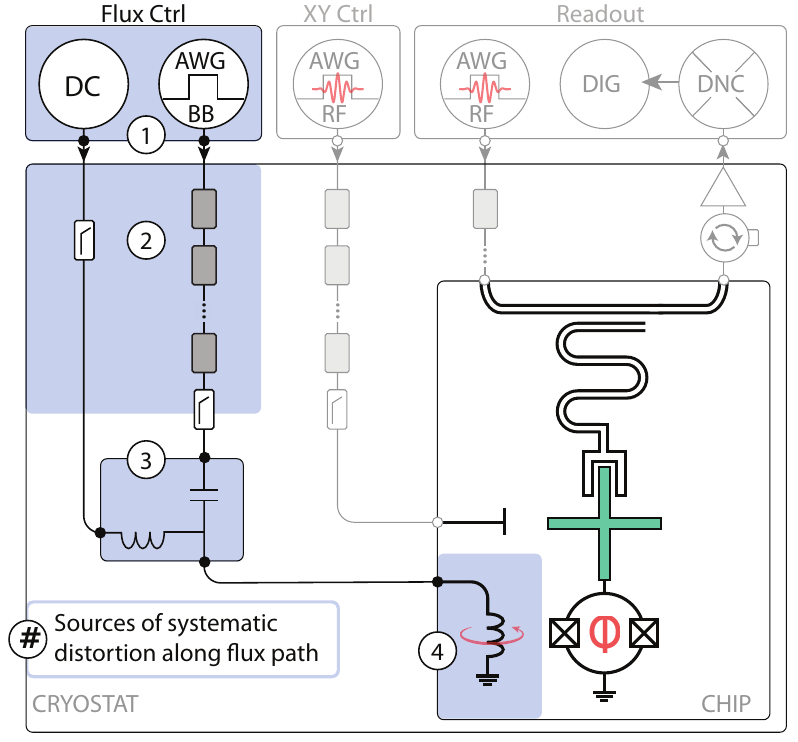}
    \caption{Concept level block diagram of the flux control signal chain. Key distortion sources include: (1) AWG pulse generation, (2) cryogenic flux line components and coaxial cables, (3) bias-tee high-pass filtering, and (4) quantum chip voltage-to-flux transfer function.}
    \vspace{-10pt}
    \label{fig:signal chain}
\end{figure}

In state-of-the-art tunable superconducting qubit systems, flux-control signals are generated by an arbitrary waveform generator (AWG) operating at room temperature. This signal travels along various non-ideal electrical components, acquiring dynamical distortions (DDs) before finally reaching the transmon qubit \cite{Zurich}. The contributing distortion effects are highlighted in Fig. \ref{fig:signal chain}, namely, the AWG pulse generation at room-temperature, the flux-line components, the bias-tee, and the quantum chip response. If these DDs are not compensated for, they are detrimental to qubit gate fidelity and repeatability. As an alternative to distortion compensation, it is possible to introduce a settling time on the timescale of the distortion transient effect. However, this introduces a delay that, depending on the time constants involved, can dramatically increase the experiment execution time. Instead, our approach deploys digital predistortion techniques to compensate for the different types of DDs to achieve both improved gate fidelity and optimization of the execution time of quantum algorithms. The compensation coefficients are calculated from characterization measurements of both classical and quantum distortion contributions.

\subsection{Classical Distortion Effects}
\begin{table}[t]
    \centering
    \begin{tabular}{lll} 
        \toprule
        Dist. Type      & Model & Time Constants \\
        \midrule
        AWG        & $H(s) = \frac{\omega_n^2}{s^2+2\zeta\omega_ns + \omega_n^2}$ & $[1\,\text{ns}, 100\,\text{ns}]$ \\
        Cryo Flux Line & $y(t) = (1+Ae^{-t/\tau})\cdot u(t)$ & $[1\,\text{ns}, 10\,\mu\text{s}]$ \\
        Bias-tee       & $y(t) = e^{-t/\tau}$ & $[10\, \mu\text{s}, 10\,\text{ms}]$ \\
        \bottomrule
    \end{tabular}
    \vspace{5pt}
    \caption{Distortion models for the different classical distortion types considered in this paper along with their respective time constants. $s$ is the Laplace-domain variable; $\omega_n$ is the natural angular frequency of the system; $\zeta$ is the damping coefficient; $\tau$ is the time constant; $A$ is the amplitude coefficient.}
    \vspace{-15pt}
    \label{table:dist_effs}
\end{table}
Classical distortion effects are generated by the room-temperature and cryogenic flux line components leading down to the QPU. The distortion effects considered in this paper are listed in Table \ref{table:dist_effs} along with their typical range of time constants.
The AWG pulse is modeled using an under-damped, second-order system based on instrument characterization measurements \cite{FFT_PhD}. Additionally, the combination of the components of the cryogenic flux line can lead to an exponential under/over-shoot distortion effect \cite{Zurich}.  Finally, the bias-tee model is modeled using an exponential decay response akin to a high-pass filter \cite{Zurich}.

\subsection{Quantum Chip Voltage-to-Flux Transfer Function}
The flux-control signal reaches the quantum chip to modulate the magnetic flux of the SQUID. This operation can be described by a voltage-to-flux transfer function that can be identified \textit{in-situ} if we use the qubit as a sensor \cite{S_Gustav}, \cite{Cryoscope}, \cite{MITpaper}. The state of a superconducting qubit can be measured indirectly through the resonator using the dispersive shift readout technique \cite{QEng}. This is vastly different from the typical input-output measurements that are conducted to characterize classical components. The voltage-to-flux transfer function can be extracted from qubit-state measurements by inverting the flux-to-qubit frequency equation to derive the magnetic flux from the qubit accumulated phase \cite{Zurich}, \cite{Cryoscope}, \cite{MITpaper}. In this work, we have achieved this operation using a Ramsey-style experiment known as Cryoscope \cite{Cryoscope} that is described in section \ref{subsec: RamseyExp}. 





\section{Digital Predistortion Simulations}
\begin{figure}[t]
    \centering
    \includegraphics[width=0.85\linewidth]{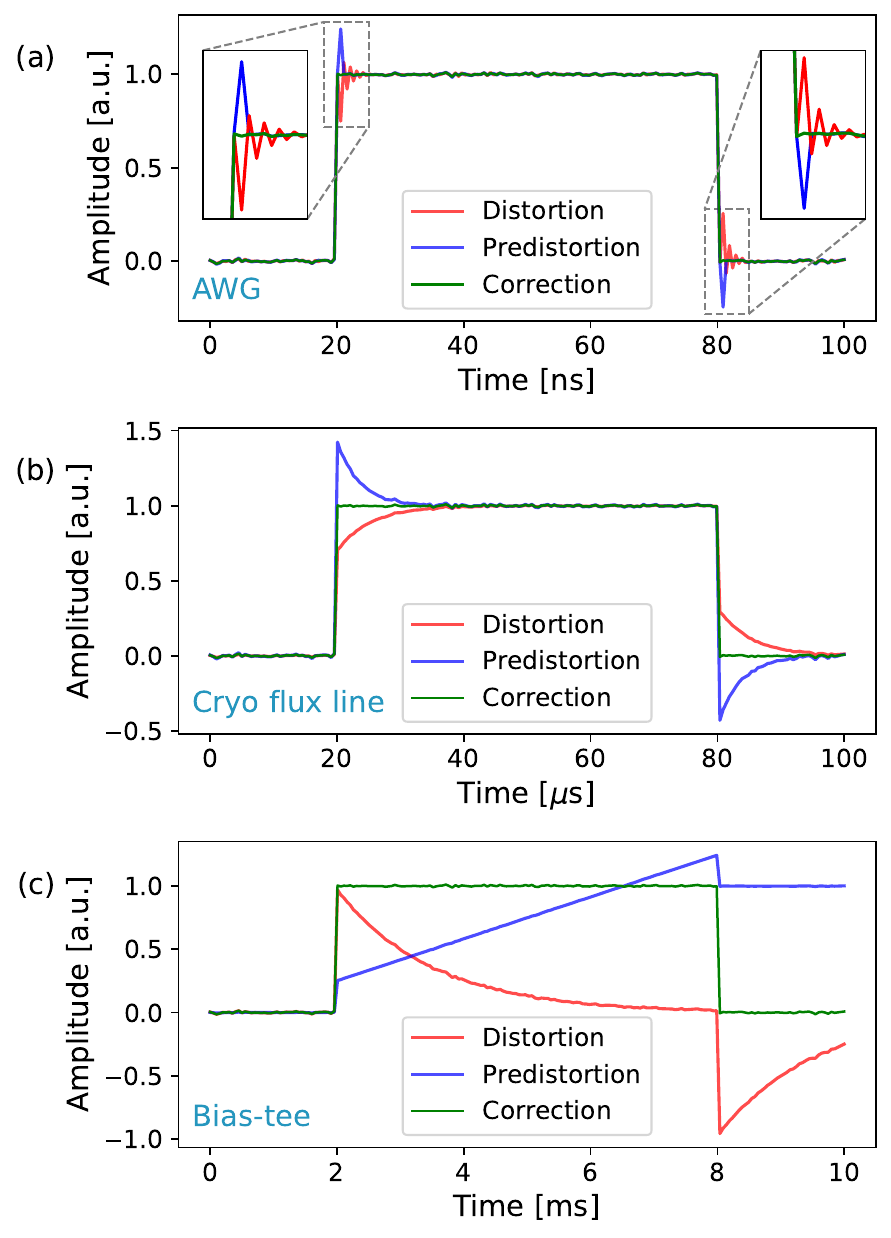}
    \caption{Simulation results showing the distorted signal, the predistorted signal, and the corrected signal as obtained modeling distortion effects coming from (a) an AWG, (b) the cryo flux line, and (c) a bias-tee.}
    \vspace{-10pt}
    \label{fig: dpd sim corrections}
\end{figure}

We created a simulation environment to test the application and compensation of different distortion effects on the flux-control line illustrated in Fig. \ref{fig:signal chain}. The distortion types are modeled through the relevant difference equation. To compensate for the modeled distortion effects, an IIR filter with feedforward taps $b_i$ and feedback taps $a_j$ is applied to the input signal $x[n]$ to obtain a predistorted signal $y[n]$ that can be described as
\begin{equation} \label{eq: IIR}
    y[n] = \sum_{i=0}^{M_b}b_ix[n-i] + \sum_{j=1}^{M_a}a_jy[n-j],
\end{equation}
where $M_b$ and $M_a$ are, respectively, the number of feedforward and feedback taps used. To optimize usage of the field programmable gate array (FPGA) resources, the number of taps $M_b$ and $M_a$ should be minimized. The filter coefficient values for a successful compensation are calculated using a linear least-squares (LS) algorithm approach \cite{FFT_PhD}. Fig. \ref{fig: dpd sim corrections} shows the successful correction for the classical distortion types considered. The optimum number of feedforward and feedback taps were then determined by using the criteria of minimum normalized mean-squared error (NMSE) between the corrected output and the original input signal. We have quantified the successful compensation of different distortion effects using the $\log(\text{NMSE})$ \cite{FFT_PhD}. Simulation results (a)-(c) showed a $\log(\text{NMSE})$ < -30$\,$dB with a minimum number of filter coefficients $M_a\!=\!1, M_b\!=\!2$. As a next step, we have measured a quantum device to test if the digital predistortion (DPD) technique based on Eq. (\ref{eq: IIR}) could effectively compensate for both classical and quantum DDs measured altogether.


\section{QPU Measurement Results}
\begin{figure}[t]
    \centering
    \includegraphics[width=0.85\linewidth]{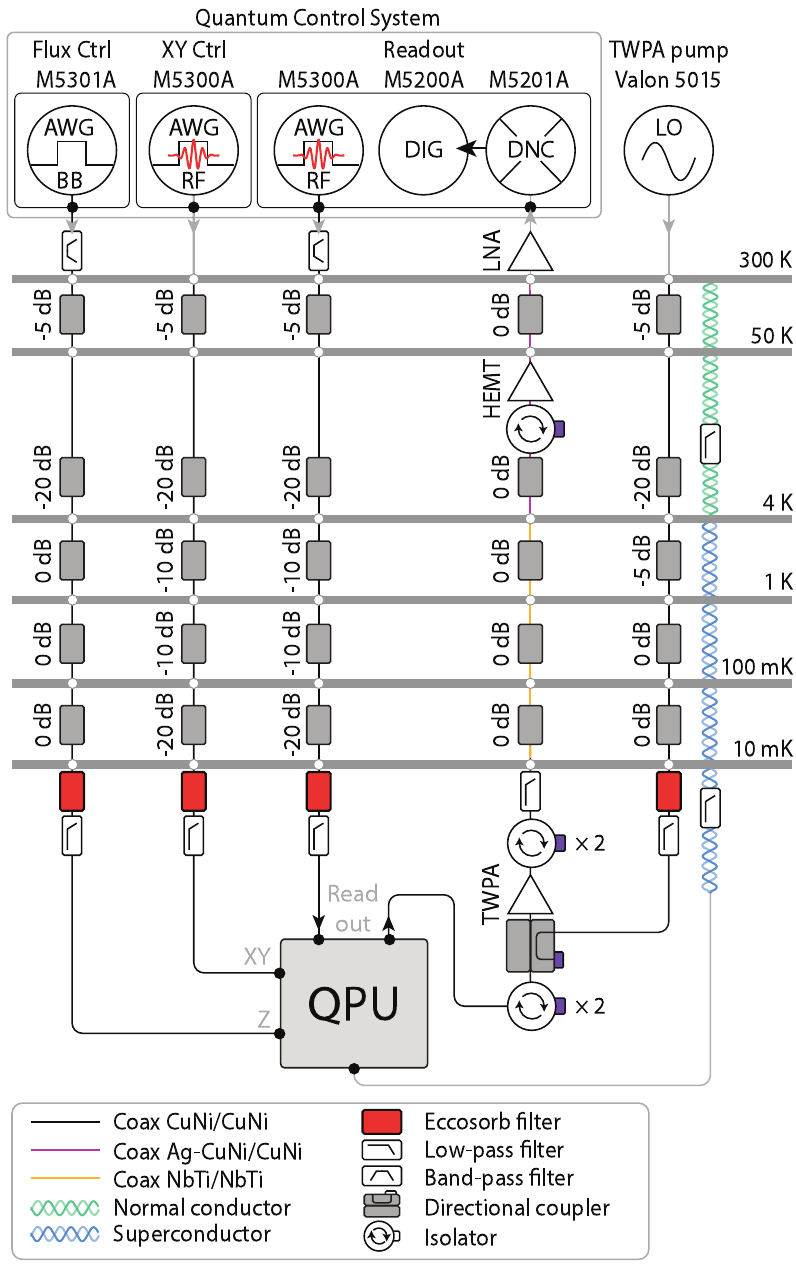}
    \caption{Measurement setup used for the quantum experiments described in this paper. The gray separators represent the different temperature stages inside the dilution refrigerator setup inside the lab of the Center for Quantum Technologies (CQT) at the National University of Singapore (NUS).}
    \vspace{-10pt}
    \label{fig:fridge wiring}
\end{figure}
Fig. \ref{fig:fridge wiring} shows a schematic of the cryostat setup that was used for the experiments. The following results were obtained from the measurements of qubit Q1 in a 20-qubit IQM Garnet QPU \cite{IQM_QPU}. Notably, an advantage of the Keysight Quantum Control System (QCS) is that the BB AWG (M5301A) can directly generate composite DC+BB pulses \cite{KeysightQCS}, thus mitigating the need for a bias-tee. Therefore, we can disregard the bias-tee distortions up to $10\text{ ms}$ timescales that are beyond the qubit coherence times. In addition, the AWG generates microwave signals using direct digital synthesis (DDS), meaning we do not need to model and correct for the distortion effects resulting from up-conversion using analog mixers. As a result, in our setup we can infer flux line distortion effects by performing the QPU characterization measurements described in this section. To perform this characterization, we first need to measure the dependence of the qubit frequency on applied magnetic flux. This is performed via a 2D qubit spectroscopy experiment in which the voltage applied on the $Z$ line is varied. This allows us to extract the voltage-to-flux transfer function and then compensate for it using DPD. Following this, the qubit is probed using a Ramsey-style experiment that allows for the measurement of the qubit's accumulated phase as a function of the flux-control voltage \cite{Cryoscope}. As the flux-to-qubit frequency is approximately a quadratic relationship, we therefore linearize the transfer function up to the voltage-to-flux control.

\subsection{2D Flux Spectroscopy}
\begin{figure}[t]
\vspace{-10pt}
    \centering
    \includegraphics[width=1.02\linewidth]{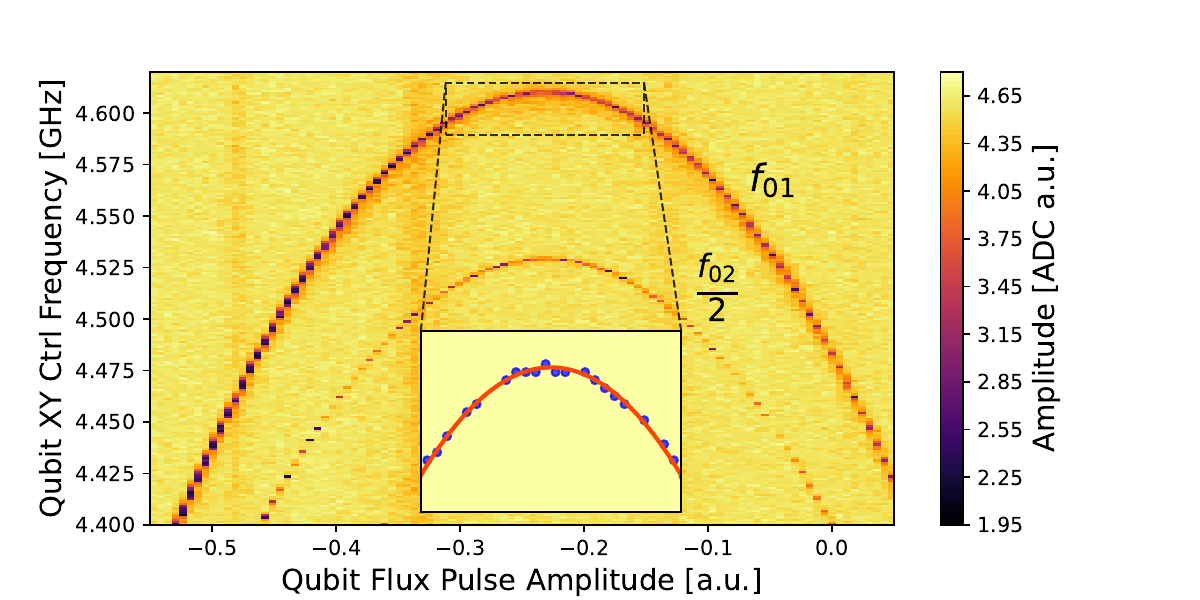}
    \caption{2D-flux spectroscopy experiment results shown as a heat map highlighting the sharp change in transmission. The upper bold curve and lower faint curve correspond to the $f_{01}$ and $f_{02}/2$ transitions, respectively. The zoomed-in inset shows the extracted resonant frequencies (blue dots) that are used to fit to the flux-frequency equation (orange line).}
    \vspace{-10pt}
    \label{fig:2D Flux Spec}
\end{figure}

For the symmetric transmon qubit, the frequency is related to the magnetic flux through its SQUID loop via the relation \cite{QEng}
\begin{equation} \label{eq: freq-flux}
    f_q(\Phi(t)) = \frac{1}{h}\left(\sqrt{8E_CE_J\left|\cos\left(\pi\frac{\Phi(t)}{\Phi_0}\right)\right|}-E_C\right),
\end{equation}
where $h$ is Planck's constant, $E_C$ and $E_J$ are the charging and Josephson energies, respectively, and $\Phi_0\!=\!h/2e\!=\!2.07\cdot 10^{-15}\,\text{Wb}$ is the superconducting magnetic flux quantum with $e$ as the electron charge. The aim is to extract $f_q(t)$ and algebraically invert Eq. (\ref{eq: freq-flux}) to reconstruct the qubit's magnetic flux response $\Phi(t)$. We perform a spectroscopy experiment by playing a long RF pulse on the $XY$ line and varying its frequency. When the frequency matches that of the qubit, there will be a sharp change in the transmission around the frequency of the readout resonator. To measure the flux dependence of the qubit frequency as predicted by Eq. (\ref{eq: freq-flux}), we simultaneously apply a 2.4 GSa/s AWG pulse of the same duration on the $Z$ line and vary its amplitude, giving the results in Fig. \ref{fig:2D Flux Spec}. The resulting data is then fitted to Eq. (\ref{eq: freq-flux}) and the relevant constants are extracted. The pulse durations are fixed to be longer than the distortion timescales, which do not exceed 100 ns in our setup, and a DC offset is played to cancel any flux offsets, hence why the maxima do not appear at $\Phi\!=\!0$. 

\subsection{Ramsey-Style Experiment}
\label{subsec: RamseyExp}
\begin{figure}[t]
    \centering
    \centerline{\includegraphics[width=0.95\linewidth]{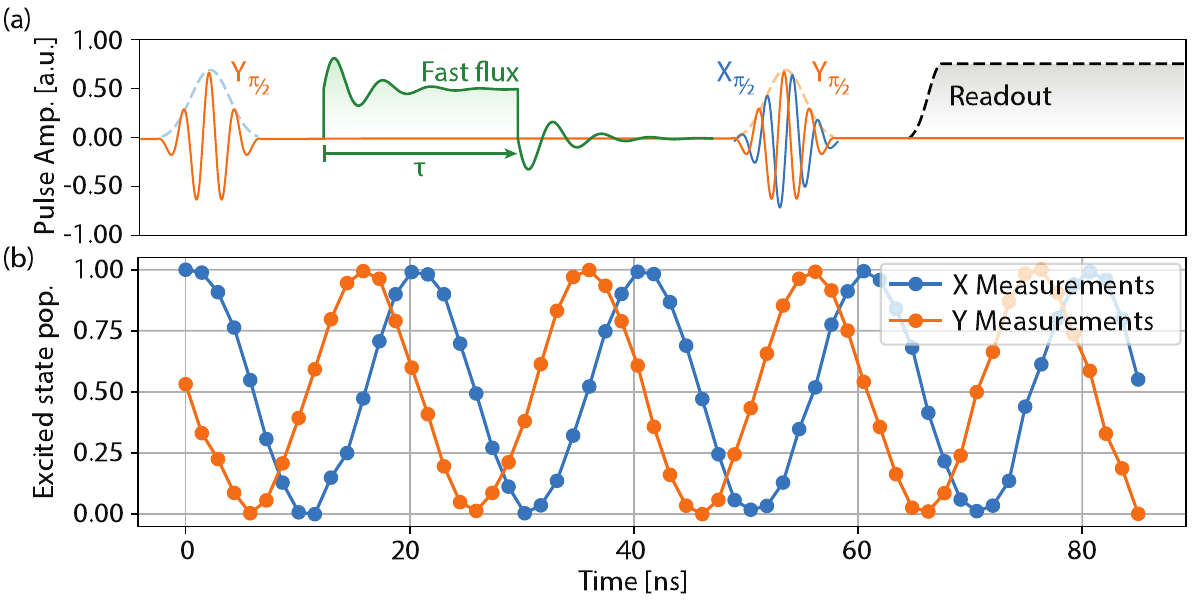}}
    \caption{(a) Pulse sequence depicting the Ramsey-style experiment. (b) Qubit-state measurements from the Ramsey-style experiment encoding the accumulated phase that enables the extraction of the voltage-to-flux transfer function.}
    \vspace{-10pt}
    \label{fig: XYmeas}
\end{figure}

After extracting the flux-frequency relation from spectroscopy, we conduct a Ramsey-style experiment that begins with the qubit prepared in the ground state $|0\rangle$. By applying a $Y_{\pi/2}$ pulse to the XY line, we drive the qubit into an equal superposition state $|+\rangle\!=\!(|0\rangle + |1\rangle) / \sqrt{2}$, corresponding to the equator of the Bloch sphere \cite{QEng}. We then apply a flux-control step pulse $V_\tau(t)$ to the flux line, with variable duration, $\tau$. The application of this pulse changes the qubit frequency by a detuning that is proportional to the amplitude of $V_\tau(t)$. The amount of phase rotation is instead proportional to  $\tau$. At this point, we rotate the qubit's Bloch sphere vector from its superposition state onto a measurement basis by applying another $X_{\pi/2}$ or $Y_{\pi/2}$ pulse on the qubit XY control line. Finally, the state of the qubit is read out by probing the readout resonator. This pulse sequence is illustrated by Fig. \ref{fig: XYmeas}(a). We repeat this experiment, varying the duration of the flux-control pulse, $\tau$. In this experiment, the range for sweeping the pulse duration $\tau$ is limited by the qubit coherence times which in our case were measured to be $T_1\!=\!21.5\,\mu\text{s},\,T_2^*\!=\!4.9\,\mu\text{s}$ using the energy relaxation and Ramsey experiments, respectively. 
Fig. \ref{fig: XYmeas}(b) shows the qubit excited state population, after being normalized to be between 0 and 1, for when an $X_{\pi/2}$ or a $Y_{\pi/2}$ pulse is applied. Each measurement point is obtained for a longer duration $\Delta\tau$ that increases the accumulated phase accordingly, leading to the oscillating pattern in Fig. \ref{fig: XYmeas}(b). From these measurements the accumulated phase is then extracted, unwrapped, and differentiated to find the instantaneous frequency detuning for each time step. This is finally used to invert Eq. (\ref{eq: freq-flux}) and reconstruct the voltage-to-flux response. 

\subsection{DPD Compensation of Distortion Effects}
\begin{figure}
    \centering
    \includegraphics[width=0.95\linewidth]{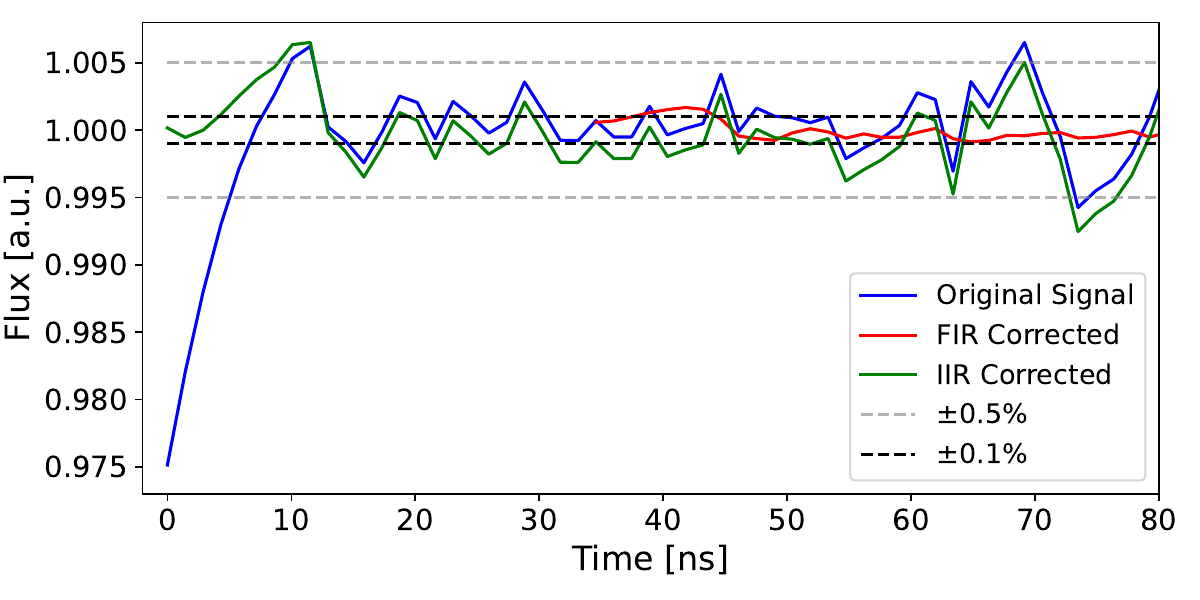}
    \caption{The voltage-to-flux response of the transmon qubit characterized using Cryoscope (in blue) as well as the IIR correction (in green) and the FIR correction (in red) once the FIR filter reached full memory.}
    \vspace{-10pt}
    \label{fig: DPD results}
\end{figure}

Once we have reconstructed the voltage-to-flux transfer function, shown by the blue curve in Fig. \ref{fig: DPD results} (after normalizing by dividing by the mean), we can use it to calculate the DPD coefficients that can compensate for it. 
In our setup, without bias-tees and other sources of distortion effects with time constants greater than the qubit coherence times, we can assume that the measured voltage-to-flux transfer function also accounts for all the other distortion effects accumulated by the flux-control signal along the flux line. In Fig. \ref{fig: DPD results} we can observe that the bulk of the distortion occurs in the first 20$\,$ns. By using an IIR filter with one feedback and two feedforward taps, we could successfully compensate for the initial rise-up transient and keep deviations within 0.65\% from the ideal shown by the green curve in Fig. \ref{fig: DPD results}. The addition of a second level of correction using an FIR filter (after reaching filter depth) achieves deviations within the $0.17\%$ mark, which is shown by the red curve in Fig. \ref{fig: DPD results}. This demonstrates that a two-stage correction deploying an IIR and a FIR filter can successfully correct for all the combined distortion effects along the flux line, including the contributions coming from the quantum chip. 

\section{Conclusion}
In this work, we have presented a DPD technique to linearize the flux-control of flux-tunable superconducting qubits. The QPU measurement results show that the demonstrated technique successfully compensates for flux-line distortion effects; an IIR filter suppresses deviations to 0.65\% from the ideal linear behavior, while a subsequent FIR filter achieves a 0.17\% margin. We have also identified the minimum number of feedback and feedforward taps that are necessary to compensate for the observed distortion effects associated with flux-control lines, making this optimization procedure ideal for integration into programmable hardware of a quantum control system. Future work will extend this DPD technique to tunable couplers and multi-qubit architectures, to facilitate the path toward high-fidelity gates in large-scale QPUs.




\bibliographystyle{IEEEtran}
\bibliography{IEEEexample}

\end{document}